# Science Under Attack

But **Nicholas Maxwell** argues it's the wrong war

**The Philosopher's Magazine** Issue 31, 3rd Quarter 2005, pp. 37-41.

**Abstract**
Some attack scientific rationality, others defend it, but both miss the point.  What both parties take to be scientific rationality is actually a species of *irrationality* masquerading as scientific rationality.  The current orthodox conception of science, taken for granted by scientists and non-scientists alike, is irrational because it suppresses problematic assumptions, inherent in the aims of science, having to do with metaphysics, values, and political and social issues.  We urgently need a more rigorous conception of science to be adopted and implemented that honestly acknowledges the problematic character of the aims of science, and seeks to improve them as science proceeds.

Science has long been under attack, at least since the Romantic movement.  William Blake objected to "Single vision & Newton's sleep" and declared that "Art is the Tree of Life... Science is the Tree of Death".  Keats lamented that science will "clip an Angel's wings" and "unweave a rainbow".  Whereas the Enlightenment had valued science and reason as tools for the liberation of humanity, Romanticism found science and reason oppressive and destructive, and instead valued art, imagination, inspiration, individual genius, emotional and motivational honesty rather than careful attention to objective fact.

Much subsequent opposition to science stems from, or echoes, the Romantic opposition of Blake, Wordsworth, Keats and many others.  There is the movement Isaiah Berlin has described as the "Counter-Enlightenment" (see chapter one of his book *Against the Current*).  There is existentialism, with its denunciation of the tyranny of reason, its passionate affirmation of the value and centrality of irrationality in human life, from Dostoevsky, Kierkegaard and Nietzsche to Heidegger and Sartre. There is the attack on Enlightenment ideals concerning science and reason undertaken by the Frankfurt school, by postmodernists and others, from Horkheimer and Adorno to Lyotard, Foucault, Habermas, Derrida, MacIntyre and Rorty (see for example Gascardi's *Consequences of Enlightenment*).  The soul-destroying consequences of valuing science and reason too highly is a persistent theme in literature: it is to be found in the works of writers such as D.H. Lawrence, Doris Lessing, Max Frisch, Y. Zamyatin.  There is persistent opposition to modern science and technology, and to scientific rationality, often associated with the Romantic wing of the green movement, and given expression in such popular books as Marcuse's *One Dimensional Man*, Roszak's *Where the Wasteland Ends*, Berman's *The Reenchantment of the World* and Appleyard's *Understanding the Present*.  There is the feminist critique of science and conceptions of science: see, for example, Fox Keller's *Reflection on Gender and Science* and Harding's *The Feminist Question in Science*.  And there are the corrosive implications of the so-called "strong programme" in the sociology of knowledge, and of the work of social constructivist historians of science, which depict scientific knowledge as a belief system alongside many other such conflicting systems, having no more right to claim to constitute knowledge of the truth than these rivals, the scientific view of the world being no more than an elaborate myth, a social construct: see Bloor, *Knowledge and Social Imagery*; Barnes, Bloor and Henry, *Scientific Knowledge: A Sociological Analysis*; Shapin and Schaffer, *Leviathan and the Airpump*; Shapin, *A Social History of Truth*; Pickering, *Constructing Quarks*; Latour, *Science in Action*.

This latter literature has provoked a counter-attack by scientists, historians and philosophers of science seeking to defend science and traditional conceptions of scientific rationality: see Gross and  Levitt, *Higher Superstition: The Academic Left and It  Quarrels with Science*; Gross, Levitt and Lewis, *The Flight from Science and Reason*, Koertge,  *A House Built on Sand*; and Brown,

*Who Rules in Science?*.

This debate between critics and defenders of science came abruptly to public attention with the publication of Alan Sokal's brilliant hoax article 'Transgressing the boundaries' in a special issue of the cultural studies journal *Social Text* in 1996 entitled *Science Wars*: see Sokal and Bricmont, *Intellectual Impostures*.

But both sides in this "science wars" debate miss the point. Those who attack scientific rationality, and those who defend it, are actually busily attacking and defending, not scientific *rationality* at all, but a species of *irrationality* masquerading as scientific rationality. Instead of fighting over the current orthodox, and *irrational* conception of science, both sides ought to turn their attention to the question of what precisely needs to be done to cure science of its current damaging irrationality, so that we may develop a kind of science that is both more rational, and of greater human value.

Science as it exists at present is irrational because it suppresses problematic assumptions, inherent in the aims of science, having to do with metaphysics, values, and political and social issues. The official intellectual aim of science is to improve knowledge of factual truth, nothing being presupposed about the truth, claims to knowledge being assessed impartially with respect to the evidence. But this is nonsense.

**Metaphysics**

It is nonsense, first, because it ignores that science must make the *metaphysical* assumption, at the outset as it were, that the universe is (more or less) comprehensible, there being explanations for phenomena to be discovered. Given any accepted scientific theory – Newtonian theory, say, or quantum theory – endlessly many rivals can easily be formulated which are just as empirically successful, if not more so. All one needs to do, to formulate such empirically successful rivals, is take the given theory and modify what it asserts about as yet unobserved phenomena, and then add on some independently testable and confirmed hypothesis. The resulting patchwork quilt theory will be more successful empirically than the given theory. If theories really were assessed impartially with respect to evidence, science would be swamped by endlessly many such patchwork quilt theories, and the whole scientific enterprise would founder.

This does not happen in practice because scientists ignore patchwork quilt theories – theories that say different things about different phenomena – whatever their empirical success might be if they were to be considered. In order to be accepted as a part of scientific knowledge, it is not enough that a theory be empirically successful; it must assert that the *same laws* apply throughout the range of phenomena to which the theory applies. The theory must be *unified* in other words. It must be *explanatory*, in the sense that the same laws predict the diverse phenomena to which the theory applies.

But in persistently accepting unified or explanatory theories only, and ignoring all disunified, non-explanatory, patchwork quilt rivals, even though these rivals are empirically more successful, science in effect makes a big, highly problematic, persistent *assumption* about the world: there is underlying unity in nature. The universe is (more or less) physically comprehensible. The universe is such that no patchwork quilt theory is true. This assumption is not itself empirically testable; it is, rather, a precondition for the process of advancing knowledge by putting testable theories to the tribunal of experimentation to get off the ground. The assumption is, in other words, *metaphysical* (untestable theories being metaphysical by definition, as it were).

Scientists don't *know* that the universe is comprehensible. This metaphysical assumption – which must be made if science is to proceed at all – is a pure article of faith. This challenges the whole orthodox conception of science, which prides itself on making no assumptions independently of evidence and being, in this respect, quite different from religion or politics. The problematic, necessary article of faith is thus *repressed*. Scientists deny that any such assumption is made, prompted by the misguided idea that they thereby preserve the rationality and the success of science, and ensure that everything in science is open to impartial empirical appraisal. But this achieves just the opposite of what is intended.

For this suppression of the big, highly problematic, metaphysical assumption that the universe is comprehensible undermines rationality, and has damaging consequences for science itself. Even if the universe is comprehensible, almost certainly it is not comprehensible in the specific way physicists assume it to be at any given stage in the development of science. Again and again, scientists have got it wrong. In the 17th and 18th centuries they assumed the universe is made up of tiny, hard corpuscles that interact by contact. In the early 19th century they assumed it is made up of point-particles that interact by means of forces at a distance. In the late 19th, early 20th centuries, they assumed it is made up of a unified field – an entity spread smoothly throughout the universe. Today, they assume it is made up of tiny strings vibrating in ten or eleven dimensions of spacetime. Tomorrow they may assume the universe is made up of tiny balls of pulsating space; or they may assume each spacetime point is made up of the entire universe as it existed before the big bang, but as a virtual state (my favourite conjecture). Who knows what wild assumption will emerge next.

The important point is that, since our specific ideas about how the universe is comprehensible are almost bound to be wrong, we need to make these untestable, *metaphysical* conjectures explicit within science, so that they can be critically assessed, so that alternatives can be considered, in the hope that in this way such assumptions can be *improved*. Quite generally, in fact, rationality requires that implicit assumptions need to be made explicit so that they can be criticized and improved. We have the following requirement for rationality:

**Rationality Principle**: Assumptions that are substantial, influential, problematic and implicit need to be made explicit so that they can be critically assessed, so that alternatives can be considered, in the hope that this will lead to the assumptions being improved.

In suppressing the substantial, highly influential and problematic metaphysical assumption of comprehensibility, in failing to make this implicit assumption explicit, scientists violate this vital *Rationality Principle*. They damage science. The orthodox view that science seeks truth, nothing being presupposed about the truth, is thus untenable, destructive and irrational. Science seeks, not truth *per se*, but rather *explanatory truth*, truth presupposed to be explanatory.

**Values**

But this is only the first hint of what is wrong with the official view of science. For science does not just seek explanatory truth. More generally, it seeks *important* truth. The search for explanatory truth is just a special case of the more general search for important truth. Science seeks to acquire knowledge deemed to be *of value*, either of value intellectually or culturally – because it enhances our understanding of the world around us or illuminates matters especially significant to us, such as our origins – or of value practically or technologically, in enabling us to achieve other ends of value, such as health, food, shelter, travel, communications. It is inevitable that values should be inherent in the aims of science. Endlessly many matters of fact are available for scientific investigation. The entire scientific community might pick on a specific matchbox, and devote centuries of research to improving our knowledge about it, its precise composition, size, number of composite atoms, history, manufacture. Inevitably, scientists must choose to pursue some lines of research and ignore infinitely many alternative possible research avenues. In thus choosing scientists in effect make decisions about what they deem to be important, what less important. Simply in order to enter the body of scientific knowledge, a result must be deemed to be sufficiently important. A scientific paper reliably reporting new factual knowledge will not be accepted for publication if judged to be too trivial. And it is not just that it is inevitable that values are inherent in the aims of science; it is *desirable* as well. We want, we need, science to discover interesting and useful knowledge. A science that accumulated a vast store of irredeemable triviality would not be judged to be making progress, and would be of no use to us at all.

But if *metaphysics* is problematic, *values* are even more profoundly problematic. Of value to whom? And in what way? Who is to decide? Scientists, under the mistaken idea that they are preserving the rationality, the objectivity, of science, deny that values have any role to play within the intellectual domain of science, and insist that science seeks knowledge of value-neutral fact. But in repudiating the role of values in this way, the scientific community, once again, achieves the exact opposite of what is intended. As before, the *Rationality Principle* is violated. The rationality, objectivity, success and human value of science are all undermined. Precisely because value assumptions, implicit in the aims of science, are substantial, influential, and profoundly problematic, they need to be made explicit so that they can be critically assessed, so that alternatives can be considered, in the hope that the values inherent in the aims and priorities of research can be improved. Failure to do this will inevitably result in scientific research which may more or less reflect the interests of the wealthy and powerful, responsible for funding science, and the interests of scientists themselves, but which will fail to reflect the needs of the majority of people on earth, the poor and powerless. This is the reality of science today.

    Writing in the seventeenth century, Robert Boyle, one of the founding fathers of modern science, had this to say about what he called the 'Invisible College' - a sort of forerunner of the Royal Society, and thus of organized scientific research. "The 'Invisible College' [consists of] persons that endeavour to put narrow-mindedness out of countenance by the practice of so extensive a charity that it reaches unto everything called man, and nothing less than an universal good-will can content it. And indeed they are so apprehensive of the want of good employment that they take the whole body of mankind for their care." A modern science and technology that put into practice the spirit of Boyle's Invisible College - thus genuinely devoting itself to the welfare of humanity - would today give priority to the problems and needs of the poorest people on earth. Problems of third-world sanitation, agriculture, malnutrition, disease, housing, transport, education, appropriate technology, would be the central focus of much of the world's scientific and technological research. But scientific and technological research as it exists today, pursued mostly in the wealthy, industrially advanced world, is devoted primarily to the interests of industry, commerce, governments, and the military, rather than those of the poor. Such a state of affairs is more or less inevitable as long as science does not possess the institutional bodies, such as "science and human need commissions", designed to correct mismatches between priorities of research and priorities of human need. In short, as long as the role of values in the intellectual domain of science is denied, and the *Rationality Principle* is violated, science will continue to fail humanity.

**Politics**

    And it goes further. Science is pursued in a social, cultural, economic and political context. It is a part of various social, economic and political projects which seek to achieve diverse human objectives. But the idea that science is an integral part of humanitarian or political enterprises with political ends clashes, once again, with the official view of science that the intellectual aim of science is to improve factual knowledge of truth *per se*.. The political objectives of science are repressed. As before, scientists, under the mistaken idea that they are preserving the rationality, the objectivity, of science, deny that politics have any role to play within the intellectual domain of science. But in repudiating the role of politics in this way, the scientific community, once again, achieves the exact opposite of what is intended. The *Rationality Principle* is violated and the human value of science is undermined. For, of course, the political objectives of science, like all our political objectives, are profoundly problematic. These need to be made explicit so that they can be scrutinized, so that alternatives can be developed and considered, and so that the humanitarian and political objectives of science can be improved. The pretence that science does not have this political dimension compromises the rationality of science, and its human value. It lays science open to becoming a part of economic, corporate and political enterprises that are not in the best interests of humanity.

    It is not just natural science that has problematic aims because of implicit assumptions concerning metaphysics, values and social and political objectives. This is true of social inquiry and the humanities as well. Indeed, the above argument applies to the whole academic enterprise. Instead of clinging to the official, and absurd, idea that the proper intellectual aim of academia is to acquire knowledge of truth *per se*, we need, rather, to acknowledge that the real, and profoundly

problematic aims of academia are social, political and humanitarian in character.  These social, humanitarian aims need to be made explicit so that they can be critically assessed, so that alternatives can be considered, and so that they can be improved.  This needs to be done in the interests of reason, and the human value of inquiry.  As I argue in my book *Is Science Neurotic?*, the proper basic aim for academic inquiry is to help humanity learn a bit more *wisdom* – wisdom being the capacity to realize what is of value in life, for oneself and others, wisdom including knowledge and technological know-how, but much else besides.

The upshot of the line of argument just indicated is that we need to bring about a revolution in the aims and methods of science, and of academic inquiry more generally.  Natural science needs to change; its relationship with the rest of academic inquiry, with social inquiry and the humanities, needs to change; and most importantly and dramatically, academic inquiry as a whole needs to change.  The basic task of the academic enterprise needs to become to help humanity learn how to tackle its problems of living in more cooperatively rational ways than at present.  We need to put the intellectual tasks of articulating our problems of living, and proposing and critically assessing possible solutions, possible and actual *actions*, at the heart of academic inquiry. We need a kind of academic inquiry that acts as a people's civil service, doing openly for the public what actual civil services are supposed to do in secret for governments.

Natural science, despite its flaws, has massively increased our knowledge and technological know-how.  This in turn has led to a massive and sometimes terrifying increase in the power of some to act.  Often this unprecedented power to act is used for human good, as in medicine or agriculture.  But it is also used to cause harm, whether unintentionally (initially at least) as when industrialization and modern agriculture lead to global warming, destruction of natural habitats and rapid extinction of species, or intentionally, as when the technology of war is used by governments and terrorists to maim and kill. Before the advent of modern science, when we lacked the means to do too much damage to ourselves and the planet, lack of wisdom did not matter too much.  Now, with our unprecedented powers to act, bequeathed to us by science, lack of wisdom has become a menace.  This is the crisis behind all the other current global crises: science without wisdom.  In these circumstances, to continue to pursue knowledge and technological know-how *dissociated* from a more fundamental quest for wisdom can only deepen the crisis.  As a matter of urgency, we need to bring about a revolution in the academic enterprise so that the basic aim becomes to promote wisdom by intellectual and educational means.  At present science and the humanities betray both reason and humanity.

The current sterile debate between those who attack and defend scientific rationality urgently needs to be transformed.  Both parties need to recognize that what is at present being fought over is a damaging kind of *irrationality* masquerading as scientific rationality.  We need *more* scientific rationality, not less – more *authentic* scientific rationality.  Above all, we need to develop a kind of academic inquiry rationally devoted to promoting wisdom, to helping humanity learn how to make progress towards a genuinely civilized world.  In our rapidly changing world, interconnected, overcrowded, menaced by modern armaments and the consequences of global warming, fraught with poverty, war and injustice, wisdom and civilization have become, not luxuries, but necessities.

**Nicholas Maxwell is the author of *Is Science Neurotic?* (Imperial College Press), *Understanding Scientific Progress* (Paragon House) and *In Praise of Natural Philosophy* (McGill-Queen's University Press).**